\documentclass[conference]{IEEEtran}
\IEEEoverridecommandlockouts
\usepackage{cite}
\usepackage{amsmath,amssymb,amsfonts}
\usepackage{algorithmic}
\usepackage{graphicx}
\usepackage{textcomp}

\usepackage{tikz}
\usepackage{hyperref}
\usepackage{flushend}
\usepackage{orcidlink}

\usepackage{amsmath,amssymb,amstext} 

\usepackage[ruled,linesnumbered]{algorithm2e}
\usepackage{listings, xcolor}

\usepackage{xcolor}
\def\BibTeX{{\rm B\kern-.05em{\sc i\kern-.025em b}\kern-.08em
    T\kern-.1667em\lower.7ex\hbox{E}\kern-.125emX}}
    
\newcommand{\tool}{Gas Gauge}
\newcommand{\dete}{Detection Phase}
\newcommand{\iden}{Identification Phase}
\newcommand{\core}{Correction Phase}
\newcommand{\OoG}{Out-of-Gas}
\newcommand{\numcontracts}{1,000}

\begin{document}

\title{\tool: A Security Analysis Tool for Smart Contract \OoG\ Vulnerabilities}

\author{
    {Behkish Nassirzadeh \orcidlink{0000-0002-3227-1409}, Vijay Ganesh \orcidlink{0000-0002-6029-2047}}\\
    University of Waterloo\\
    Waterloo, Canada\\
    \{bnassirz, vijay.ganesh\}@uwaterloo.ca
  \and
    Huaiying Sun \orcidlink{0000-0003-3693-6743}\\
    East China University of Science and Technology\\
    Shanghai, China\\
    ecustshy@foxmail.com
    \and
    Sebastian Banescu \orcidlink{0000-0003-0771-4826} \\
    Quantstamp\\
    Munich, Germany\\
    sebi@quantstamp.com
}

\maketitle

\begin{abstract}
In recent years, we have witnessed a dramatic increase in the adoption and application of smart contracts in a variety of contexts. However, security vulnerabilities pose a significant challenge to the continued adoption of smart contracts. An important and pervasive class of security vulnerabilities that afflicts Ethereum smart contracts is the {\it gas limit DoS on a contract via unbounded operations}. These vulnerabilities result in a failed transaction with an \texttt{"out-of-gas"} error and are often present in contracts containing loops whose bounds are affected by end-user input. To address this issue, we present \tool, a tool aimed at detecting \OoG\ DoS vulnerabilities in Ethereum smart contracts. The \tool\ tool has three major components: The \dete, \iden, and \core. The \dete\ component consists of an accurate static analysis approach that finds and summarizes all the loops in a smart contract. The \iden\ component uses a white-box fuzzing approach to generate a set of inputs that causes the contract to run out of gas. Lastly, the \core\ component uses static analysis and run-time verification to predict the maximum loop bounds consistent with allowable gas usage and suggest appropriate repairs to the tool's users. Each part of \tool\ can be used separately or all together to detect, identify and help repair contracts vulnerable to \OoG\ DoS vulnerabilities. \tool\ was tested on \numcontracts\ real-world solidity smart contracts. When compared to seven state-of-the-art tools, we show that \tool\ is the most effective (i.e., has no false positives and false negatives) while being competitive in terms of efficiency.
\end{abstract}


\maketitle

\section{Introduction}
\label{intro}
Smart contracts are one of the main applications of leading blockchains such as Ethereum~\cite{eth}. Smart contracts are executable programs that allow building a programmable value exchange or a contract between various parties without the need for a trusted third-party. While smart contracts bring many benefits to the blockchain ecosystem, they suffer from various security vulnerabilities. Certain vulnerabilities can be exploited by attackers to steal funds from a smart contract, while others can cause funds to be locked indefinitely. One security issue plaguing Ethereum smart contracts is \OoG\ Denial of Service (DoS) vulnerabilities. Every operation in an Ethereum smart contract costs a certain amount of gas, a measurement unit for the amount of computational effort required to execute said operation or transaction, paid by the transaction initiation party. Each block comes equipped with an upper bound on the amount of gas that can be spent to compute all the transactions within that block. This is called the \emph{Block Gas Limit}. Since a transaction cannot exceed one block, the transaction gas limit is also bound by the block gas limit ~\cite{ethereum}. One of the kinds of gas-related vulnerabilities is \emph{DoS with Unbounded Operations}, also known as \emph{Unbounded Mass operations }~\cite{Madmax}. In this case, the execution of the transaction requires more gas than the block gas limit. As a result, The execution of one or more functions in a smart contract vulnerable to \OoG\ can be blocked indefinitely if \OoG\ conditions are not appropriately handled. This is the type of vulnerability that we focus on in this paper. 

Many Ethereum wallets, such as Metamask~\cite{MetaMask}, have a built-in mechanism to estimate the cost of a transaction statically, right before it is executed. However, there are certain cases when the gas estimation is incorrect or impossible to estimate. For example, if multiple operations are performed inside a loop traversing a dynamic array or mapping~\cite{Stackexchange}. The Metamask Support website acknowledges this issue~\cite{MetaSupport} and indicates that users should manually adjust the transaction gas limit according to one of the latest passing transactions for the smart contract function they are trying to call.

In order to address the above-mentioned problem, we present a tool, \tool, that automatically detects and suggests remediation  for \OoG\ vulnerabilities in Ethereum smart contracts. We were motivated by two factors in designing \tool. First, to-date, the most widely used method for detecting and repairing \OoG\ vulnerabilities is manual security audits which cannot continue to scale with the ever-growing size, complexity, and the number of smart contracts. Second, as we show in this paper, existing automated methods based on static, symbolic, or run-time analysis are plagued by either high false negative rates or scalability issues. Our insight is that the best way to address this problem is to use an appropriate combination of static and dynamic analysis methods. Hybrid methods can outperform all other approaches because, in order to detect these vulnerabilities, one needs to determine the loops or functions that can lead to an \OoG\ DoS (best done via static analysis), and then perform gas analysis to determine the exact point, e.g., loop iteration, when \OoG\ occurs (best done using appropriate dynamic analysis).

\subsection{Contributions.}
The contributions we make in this paper are as follows:
\begin{enumerate}
\renewcommand{\theenumi}{\roman{enumi}}
    \item Design and implementation of a static analysis and run-time verification tool, \tool \footnote{\url{https://gasgauge.github.io/}}, aimed at automatically detecting \OoG\ DoS vulnerabilities in Ethereum smart contracts. We implemented three techniques in \tool\ that are critical for identifying \OoG\ DoS vulnerabilities. The first technique is a static analysis method that identifies and summarizes loops in smart contracts. The second is a white-box fuzzing technique that triggers \OoG\ errors in smart contracts that contain publicly accessible functions with loops whose bounds are influenced by inputs (To the best of our knowledge, this feature is unique to \tool). The third is a method for identifying a threshold, as a function of input and state variables, at which \OoG\ errors can be triggered in a contract-under-analysis (To the best of our knowledge, this feature is also unique to \tool).
    \item An extensive  evaluation of the \tool\ against seven state-of-the-art tools:  GASTAP/Gasol ~\cite{gasol, Gastap}, Madmax~\cite{Madmax}, MPro~\cite{MPro}, Mythril~\cite{Mythril}, Securify 2.0~\cite{securify2}, Slither~\cite{Slither}, and SmartCheck~\cite{smartcheck} on a benchmark of \numcontracts\ real-world smart contracts. Our evaluation found that \tool\ outperforms these state-of-the-art tools. That is, \tool\ has zero false negative and false positive rates, while at the same time has a similar efficiency profile to the fastest tool in the set, SmartCheck. Further, none of these tools provide any support for repair, a feature that is essential for the industry.
	
	\item A case study on a real-world application, Airswap, a peer-to-peer trading network for Ethereum used for trading of over \$20 million/week with current ETH value. We consulted with the engineers who manually audited Airswap and were informed that constructing the repair for the \OoG\ DoS vulnerability in Airswap took them several man-hours of work, while our \tool\ tool managed to automatically accomplish the same task on this real-world smart contract in under 10 minutes.
\end{enumerate}

Our experimental evaluation reveals that \tool\ is useful, accurate, and outperforms other state-of-the-art tools that can be used to detect \OoG\ vulnerabilities. We tested our tool on \numcontracts\ real-world smart contracts extracted from Etherscan\footnote{\url{https://etherscan.io/}}, one of the most popular Ethereum blockchain explorers. All these contracts were manually checked to ensure that they contain at least one loop since user-controlled loops are one of the main causes of \emph{DoS with Unbounded Operations} \cite{Madmax}. Our tool uses run-time analysis to ensure no false positives, and our static analysis method ensures a zero false negative rate on the evaluated benchmark. Also, we performed a case study that showed that \tool\ could be utilized in real-world projects written in Solidity to save hours of manual work and millions of USD.

\begin{figure*}[!t]
\centerline{\includegraphics[width= 7 in]{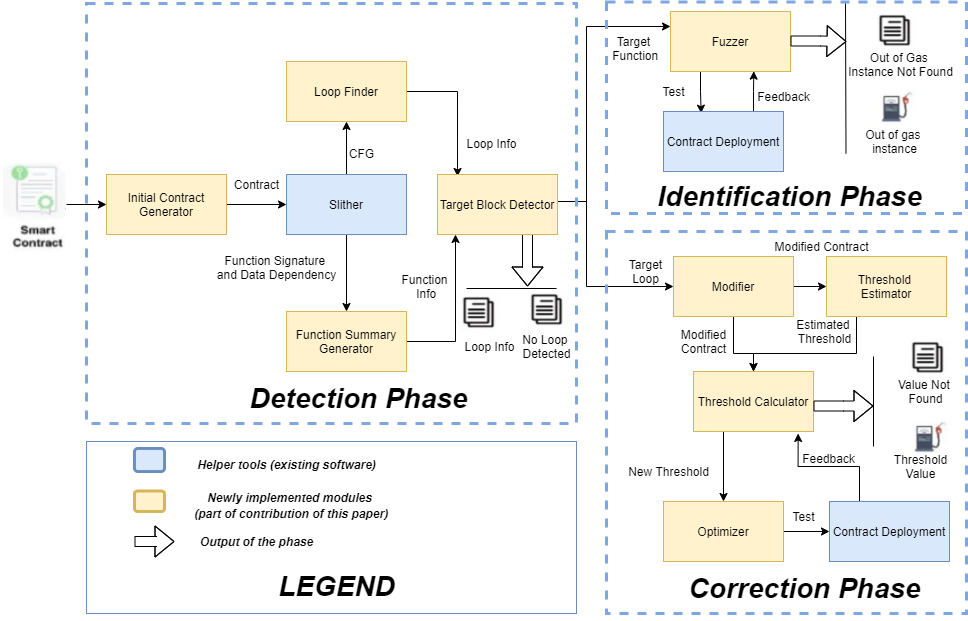}}
\centering
\caption{The Architecture of \tool}
\label{fig1}
\end{figure*}

\section{Background}
\label{background}

\subsection{Background on Smart Contracts:} Ethereum is one of the leading blockchain platforms. It is a decentralized and open-source blockchain that contains millions of accounts and billions of USD in capitalization. Hence, it is one of the most prevalent underlying technologies for smart contracts~\cite{MPro}. Smart contracts handle transactions in a cryptocurrency called Ether. They are commonly written in the Solidity language, which is a Turing-complete programming language \cite{Solidity}. It then gets compiled to Ethereum Virtual Machine (EVM) bytecode instructions to be deployed on the blockchain. Unlike traditional software programs, a smart contract is publicly accessible, transparent, and immutable. Therefore, once a smart contract is deployed on the blockchain, it cannot be altered. Thus, if errors or vulnerabilities are found in a smart contract post-deployment, they cannot be fixed a posteriori by developers unless CREATE2 is used~\cite{create2}. Several smart contract vulnerabilities have been discovered in recent years, of which \OoG\ DoS are among the common ones. A list of the most known smart contract vulnerabilities can be found on the SWC Registry website\footnote{\url{ https://swcregistry.io/}}.

\subsection{Helper Tools:} Developing an automatic gas analysis tool from Solidity source code requires a considerable implementation effort. Fortunately, many existing open-source tools make this task easier than otherwise. For example, we used the Slither ~\cite{Slither} and Truffle Suite~\cite{Truffle} as part of the \tool. Slither provides many useful APIs to collect information about a smart contract, such as data dependencies and function signatures. This is used to summarize the contract and extract the needed information for the other parts of \tool\ implementation. Also, the Control Flow Graph (CFG) generated by Slither is used to find the loops in the contract and their orders. Further, Truffle Suite is used to compile and deploy Solidity smart contracts to a test Ethereum network. This allows us to use different sets of inputs in the \iden\ of the tool and use different threshold values in the \core\ while retrieving useful gas-related information such as gas used and gas left.

\subsection{Whitebox Fuzzing:} Nowadays, security vulnerabilities in software products can be found by two fundamental methods: Code inspection of binaries and blackbox fuzzing. Blackbox fuzzing is a class of blackbox random testing that randomly mutates well-formed inputs to the program and then tests the program on the resulting data in order to trigger a bug \cite{Whitebox}. Blackbox fuzzing is an effective method to test a program; however, it can have limitations. Low code coverage is one of the leading limitations of blackbox fuzzing resulting in missing security bugs \cite{Whitebox}. An alternative approach to blackbox fuzzing is whitebox fuzzing. It is a type of automatic dynamic test generation, based on symbolic execution and constraint solving, intended for large applications' security testing \cite{Whitebox}.

\subsection{ \OoG\ Denial of Service Vulnerabilities}

\begin{figure}[h]

\begin{lstlisting}
pragma solidity >=0.4.24 <0.7;
contract SmallBank{
    address[] users;
    function addUsers(address newUser) public {
		users.push(newUser);
	}
    function addInterest(uint interest) public {
    //Heavy code to compute interest per user
    	for(uint i = 0; i < users.length; i++){
    	    users[i].call.value(interest)();
    	}
	}
}
\end{lstlisting}
\caption{Ethereum smart contract Vulnerable to DoS with Block Gas Limit}
\label{vulCode}
\end{figure}

The gas fee has to be paid by the transaction initiation party before the execution starts. Since estimating the exact gas needed can be challenging, as described in Section~\ref{intro}, the transaction initiators can specify the maximum amount of gas they are willing to pay for their transaction to be included in a block. This is known as the transaction gas limit. If the gas usage associated with a transaction surpasses this limit, the EVM raises an \OoG\ exception and aborts the associated transaction~\cite{swcregistry}. Each block has an upper bound on the amount of gas that is determined by the network and the miners. This limit is called the Block Gas Limit. A transaction cannot exceed one block, so the transaction gas limit is also bound by the block gas limit ~\cite{ethereum}. Therefore, if a transaction requires more gas than Block Gas Limit, it will revert due to \OoG, which causes the initiation party to waste gas, and no state changes occur. As a result, The execution of one or more functions in a smart contract vulnerable to \OoG\ can be blocked indefinitely if \OoG\ conditions are not appropriately handled. As a result, DoS attacks can target contracts with gas-related vulnerabilities. One of the principal kinds of gas-based vulnerabilities is \emph{DoS with Block Gas Limit} vulnerability. This vulnerability has a few different types. The most common form that mainly occurs in contracts with user-controlled loops is \emph{DoS with Unbounded Operations}. This can happen when the cost of executing a function exceeds the block gas limit~\cite{swcregistry}. This can be problematic even without any malicious intent. Generally, loops that user input determines their behavior could iterate too many times, exceeding the Block Gas Limit~\cite{Madmax}.

Figure~\ref{vulCode} demonstrates an example of a contract vulnerable to \emph{DoS with Unbounded Operations}. In this example, the number of iterations in the loop in \texttt{addInterest} is determined by the length of the variable \texttt{users}, which is controlled by the user input through \texttt{addUsers}. \texttt{addInterest} performs some expensive arithmetic calculations to compute the interest per user (not shown in the snippet) and then sends each user the interest amount. If the length of \texttt{users} is large, the computation required in the loop might reach the block gas limit, which causes the execution of the transaction to reach out of gas and revert. Thus, as the number of \texttt{users} grows, the gas needed to execute \texttt{addInterest} will increase. Ultimately, the function may become impossible to execute without raising an \OoG\ exception, at which point no user can claim their interest, and the \texttt{SmallBank} contract will suffer reputation damage and lose users.

\section{Description of \tool}
\label{components}

\tool\ is designed to address gas-based vulnerabilities of smart contracts. Since loops are the main cause of many gas-related vulnerabilities, the focus of \tool\ is on identifying and summarizing loops and then ascertaining whether they are vulnerable. \tool\ contains three major parts: the \dete, \iden, and \core. The \iden\ and \core\ require the information generated by the \dete; however, \iden\ and \core\ are independent of one another. The inputs to all methods are the Solidity source code \footnote{The contracts should be self-contained. Thus, contracts with external calls are not supported.} and the contract's gas limit (optional if only the \dete\ is used). Overall, \tool\ can detect all the loops and provide a repair to contracts vulnerable to gas-related vulnerabilities. The architecture of \tool\ is shown in Figure~\ref{fig1}.

\subsection{\dete}
The \dete\ uses a static analysis approach to efficiently and accurately detect all the loops in a smart contract. 

\subsubsection{Initial Contract Generator}
\label{initial}
The first and simplest component of the \dete\ is the Initial Contract Generator. In this step, a copy of the original contract is made. If the \core\ is needed, the copied version is formatted to make it easier for the other parts. For example, it removes all the comments and adds brackets and spaces if needed. Then, the new contract is fed to Slither. This part formats the contract in a way that does not affect the behavior but makes it easier for the static analysis section. 

\subsubsection{Target Block Detector and Its Inputs}

In this stage, inputs to the \iden\ and \core\ are generated. The input to the \iden\ is the target functions, and the input to the \core\ is the target loops. First, Slither is used to extract the contract's Control Flow Graph (CFG) and other useful information like function signatures and data dependencies. The Loop Finder uses the contract's CFG to find all the loops. Also, the information provided by Slither is used in the Function Summary Generator to summarize the contract. If only the \dete\ is needed, the program halts here and outputs the functions containing loops along with the number of loops in each function. If the \iden\ or \core\ is also needed, the types of variables affecting the loop bounds are obtained. This process utilizes a static analysis approach and uses the reports available in Slither to gather the loop conditions, the variables affecting the loops, variable dependencies, and function summaries. The variables affecting the loop bounds are classified into four groups: State, Local, Fixed, and Input variables. State variables are the contract variables whose values are persistently stored in contract storage and can be accessed by all the contract functions. Fixed variables are the ones that only carry a fixed value and are defined within the target function (in the loop bound). Input variables carry their usual meaning: the inputs of the target function. Finally, Local variables are declared and initialized inside the target function, and their context is within a function and cannot be accessed outside. If a Local variable is detected as the loop bound, the Target Block Detector performs induction to find a list of all State, Input, and Fixed variables affecting that local variable. For the white-box fuzzer, the Target Block Detector finds public functions that contain loops with bounds affected by input variables and passes them to the \iden. After identifying the target functions, the function signatures, name, and type of the input variables affecting the target loops are passed to the fuzzer. For the \core, the function signatures for all the functions containing loops, and a summary of each loop is generated. This summary includes the scope of each loop, the order of the loops if the function contains nested loops, and information about the variables affecting the loop bounds and their types. Therefore, even if a function is not passed to the \iden\ because it does not satisfy the criteria of this phase, it is still passed to the \core, and this phase finds the correct threshold values. In the \core, the contract is slightly modified, so that it can find the thresholds for any loop in any type of function.

\subsection{\iden}
In the \iden, a white-box fuzzing approach is utilized to generate a set of inputs for each user-controlled loop in a public function in a Solidity smart contract. The bounds of these loops must be affected by at least one of the input variables of the function containing the loop; otherwise, directly fuzzing the target function cannot be effective. This component takes the information from the \dete\ and the block gas limit as inputs and outputs the set of values that make the transaction go out of gas. Here, public functions mean functions that can be called from outside the contract. Thus, private functions cannot be fuzzed directly without modifying the contract. These functions are supported and checked by the \core. 

In this part, all the input variables in the target function get set to their initial values (i.e., integers are set to zero, and arrays are set to an empty array). The input values reported by the Target Block Detector get encoded in their binary representations. Then the tool picks a bit at random and flips it. Then, the binary encoding gets converted back to the original form. The only exception is arrays. In this case, the array size can generally affect the bound of a loop, so arrays are represented by 256 bits since arrays in Solidity can have up to $2^{256}$ elements ~\cite{array}. Then, the binary representation gets converted to an integer, and the array size gets set to that. If multiple input variables affect the loop bound, the binary representations get concatenated, the bit is flipped, and the concatenated value gets converted back to the original forms. Next, all the necessary files are generated automatically. Truffle Suite~\cite{Truffle} is used to deploy the contract to a test Ethereum network. A test file is generated in Solidity to call the target function with generated input values. Then, the contract is deployed, and the target function is fuzzed (tested with different input values). Suppose the test case halts and returns the remaining gas in the contract, the process repeats, and the fuzzer flips another bit. The process continues until a set of inputs is found that makes the test case abort due to an \OoG\ exception. At this point, the used set of inputs is reported as the output of the phase.  

\subsection{\core}

\begin{algorithm}[t]
\small
\caption{Estimated threshold for a loop}
\label{alg1}
\ $ \text{initial\_gas} = \text{gasleft()}$ \;
\ $ \text{iteration} = 0$ \; 
 \While{$ \text{iteration} < 2$}{
    original code inside the loop \;
    \If{$\text{iteration}  == 0$}{
     $\text{gas\_1} = \text{initial\_gas} - \text{gasleft()}$\;
     }
    $\text{gas\_2} = \text{initial\_gas} - (\text{gas\_1} + \text{gasleft()})$\;
    $\text{iteration}  \mathrel{+}=  1 $\;
    }
     $ \text{max\_iteration} =  1 + (\text{initial\_gas} - \text{gas\_1}) / \text{gas\_2}$ \;
\end{algorithm}

The \iden\ is convenient to scan the contracts before deployment and check if the contract is at risk of DoS with Block Gas Limit. Since it also provides an instance, it further helps them examine the problem. However, one of the first steps to fixing the code is to find the exact point where the \OoG\ condition starts to get triggered, and any arbitrary set of inputs is not enough. Therefore, The \core\  is designed to find the upper bound limit of the loops in a smart contract. The output is a formula based on the maximum number of allowed iterations for loops before the transaction runs out of gas. We refer to this number as the threshold of the loop.

\subsubsection{Modifier}

The Modifier makes two copies of the contract generated in the Initial Contract Generator. The first copy is to measure the gas used for each loop's first and second iteration identified by the Target Block Detector. Based on our observation, the first iteration of each loop consumes more gas than the other iterations. This is perhaps due to the gas consumption of the loop initialization, where the counter gets initialized to a starting value. The second iteration's gas usage is typically the average gas usage of all the other loop iterations. The first copy is the input to the Estimator, and the second copy is to change the loop bound to the desired value. It allows us to run each loop with a specified number of iterations and capture the gas left after that many iterations. The Threshold Calculator uses this to test different values. The modifications only have an insignificant effect on the behavior/gas usage of the contract. A public function is added as a wrapper for both modified copies to call the target function.

\subsubsection{Threshold Estimator}

The Threshold Estimator receives the first modified contract and automatically creates a Solidity test file and all the other necessary files for Truffle Suite~\cite{Truffle}. Also, to get an actual gas usage for each iteration, the new contract and test file are deployed to a test Ethereum network using Truffle Suite. The modified contract runs each loop twice and captures the gas used in each iteration. One can call the function \texttt{gasleft() returns (uint256)} that exists in the global namespace and returns to get the remaining gas at any instance~\cite{eth}. We can utilize this function to measure the gas usage of the target block of code. Next, based on the gas usage amount reported by the Truffle Suite and the gas limit, the maximum number of iterations that the loop can perform without running out of gas is estimated. Algorithm ~\ref{alg1} shows the used method method. As shown, the maximum iteration is estimated to be the amount of the initial gas before entering the loop minus the gas consumption of the first iteration divided by the average gas consumption of all the other iterations. An extra iteration is added to account for the first iteration.

\subsubsection{Threshold Calculator}

We first use the run-time/static analyzer to estimate the loop bound threshold. Thus, any value over this threshold may trigger the \OoG\ condition. Then, we used a run-time verification approach to find a more accurate value. We further use a binary search approach to cut down the action space rapidly. The action space is all the integers that can be the loop bound. The estimated threshold helps the binary search model have a proper starting point. The second set of the modified files and the Estimator's result is fed to the Threshold Calculator. The purpose is to run each loop with a specified number of iterations and capture the gas left after that many iterations. The threshold calculator uses this to test different values. The code snippets in Figure~\ref{loop1} and ~\ref{loop2} demonstrate how a contract is modified in order to find the threshold for the first loop. The code snippet in Figure~\ref{loop1} is the original contract that contains two loops and the code snippet in Figure~\ref{loop2} is the modified contract.

\begin{figure}
\begin{lstlisting}
pragma solidity >=0.4.24 <0.7;
contract TestContract{
    uint[] numbers;
    function addNumbers(uint[] calldata newNumbers) external returns(uint){
        for(uint i=0; i<newNumbers.length;i++) {
            numbers.push(newNumbers[i]);
        }
        uint sum = 0;
        for(uint j=0; j<numbers.length; j++){
            sum += numbers[j];
        }
    }
}
\end{lstlisting}

\caption{The Original Code Containing two Loops}
\label{loop1}
\end{figure}

\begin{figure}

\begin{lstlisting}
pragma solidity >=0.4.24 <0.7;
contract TestContract {
    uint[] numbers;
    function getStateVarInaddNumbers(uint[] memory input1) public returns (uint) {
        addNumbers(input1);
        return gasleft();
    } 
    function addNumbers(uint[] memory newNumbers) public returns  (uint){
        uint loopCounter = 0; 
        for(uint i=0; loopCounter < VALUE; i++) {
            numbers.push(newNumbers[i]);
            loopCounter = loopCounter + 1;
        }
        uint sum = 0;
        for(uint j=0; 1 == 0 ; j++){
            sum += numbers[j];
        }
    }
}

\end{lstlisting}

\caption{The Modified Code Containing two Loops}
\label{loop2}
\end{figure}

At this stage, by inputting different values to the target function, we can calculate different gas usages. The goal here is to find two consecutive numbers, where only the larger number makes the transaction run out of gas. Each loop is isolated to ensure the value is only affecting the target loop. The contract is deployed and run for each value, and the gas left is obtained. There is a lower bound and an upper bound limit for the search space in the Binary Search model. They are initially set to 0 and 5,000, respectively. Once the execution of a contract passes Truffle's time limit, it throws a time-out exception and stops running. Therefore, if a loop runs for too many iterations, it might trigger the time-out exception. From our experiments, if the maximum number of loop iterations is less than 5,000, the test does not trigger the time-out condition. Also, if a loop threshold is over 5,000, there is a lower chance of running out of gas as most contracts do not require that many iterations in one loop. The algorithm for this part is shown in Algorithm ~\ref{alg2}.

\begin{algorithm}[t]
\small
\caption{Run-time threshold of a loop}
\label{alg2}
 $ \text{guess} =  \text{estimated value by static and run-time analysis}$ \;
 $\text{lower} = 0$\;
 $\text{upper} = 5,000$\;
 
 \While{$\text{conditional\_expression}$}{
  $\text{gas\_left} = $deploy the contract($\text{guess}$)\;
  
  \eIf{$\text{gas\_left}$ \textless \  $ 0$}{
     $\text{new\_gas\_left} = $ deploy the contract($\text{guess} - 1$) \;
       
       \eIf{$\text{new\_gas\_left}$ \textless \  $ 0$}{
            $\text{upper} = \text{guess} - 1$\;
       }{
            $\text{threshold} = \text{guess} - 1$\;
          
           return $\text{threshold}$\;
       }
   }{
      $\text{new\_gas\_left} = $ deploy the contract($\text{guess} + 1$) \;
           
           \eIf{$\text{new\_gas\_left}$ \textgreater \   $ 0$}{
                $\text{lower} = \text{guess} + 1$\;
           }{
                $\text{threshold} = \text{guess} + 1$\;
                
                return $\text{threshold}$\;
           }
  
  }
  
  $\text{guess} = \text{upper} + \text{lower} / 2$ \;
 }
\end{algorithm}

\subsubsection{Optimizer}

The last part is the Optimizer. It has two primary purposes. First, if the lower bound reaches 5,000 or the value reported by the Estimator is over 5,000, it estimates the loop's threshold based on the original gas, gas consumed by the first iteration, gas consumed by 5,000 iterations, and the remaining gas. Secondly, to test a value in the Threshold Calculator, the system needs to run the contract with the target value, wait for the result, and then run the contract with either \texttt{value  + 1} or \texttt{value - 1}. Since each execution takes a few seconds, running the contract twice each time may take a while. Therefore, the Optimizer makes two extra copies of the generated files and simultaneously runs the three contracts with three consecutive numbers (\texttt{value}, \texttt{value  + 1}, \texttt{value  - 1}). Then, based on the feedback fit receives, it decides to use the correct extra feedback. This way, the run-time gets reduced by almost 50\%.

\begin{figure}
\begin{lstlisting}
 function hasCardExpired(uint[] calldata marketAddresses, uint numberOfTokens) public returns (bool) {
	bool _expired = false;
	for (uint i = 0; i < marketAddresses.length; i++) {
		IRealityCards rc = IRealityCards(marketAddresses[i]);
		for (uint j = 0; j < numberOfTokens; j++) {
			if (rc.cOwnerLeftDeposit(j) == 0 && rc.ownerOf(j) != address(rc)) {
				_expired = true;
			}
		}
	}
	return _expired;
}
\end{lstlisting}
\caption{The Original Code Containing a Nested Loop}
\label{nested1}
\end{figure}

\begin{figure}
\begin{lstlisting}
 function hasCardExpired(uint[] calldata marketAddresses, uint numberOfTokens) public returns (bool) {
	bool _expired = false;
	uint outerLoopCounter = 0;
	for (uint i = 0; outerLoopCounter < 1; i++) {
		IRealityCards rc = IRealityCards(marketAddresses[i]);
		uint loopCounter = 0;
		for (uint j = 0; loopCounter < VALUE; j++) {
			if (rc.cOwnerLeftDeposit(j) == 0 && rc.ownerOf(j) != address(rc)) {
				_expired = true;
			}
			loopCounter = loopCounter + 1;
		}
		outerLoopCounter = outerLoopCounter + 1;	
	}
	return _expired;
}

\end{lstlisting}
\caption{The Modified Code for the Threshold of the Inner Loop}
\label{nested2}
\end{figure}

\begin{figure}[t]
\begin{lstlisting}
 function hasCardExpired(uint[] calldata marketAddresses, uint numberOfTokens) public returns (bool) {
	bool _expired = false;
	uint outerLoopCounter = 0;
	for (uint i = 0; outerLoopCounter < VALUE; i++) {
		IRealityCards rc = IRealityCards(marketAddresses[i]);
		for (uint j = 0; 1 == 0; j++) {
			if (rc.cOwnerLeftDeposit(j) == 0 && rc.ownerOf(j) != address(rc)) {
				_expired = true;
			}
		}
		outerLoopCounter = outerLoopCounter + 1;	
	}
	return _expired;
}

\end{lstlisting}

\caption{The Modified Code for the Threshold of the Outer Loop}
\label{nested3}
\end{figure}

 \begin{figure*}[htp]
\centerline{\includegraphics[width= 6 in]{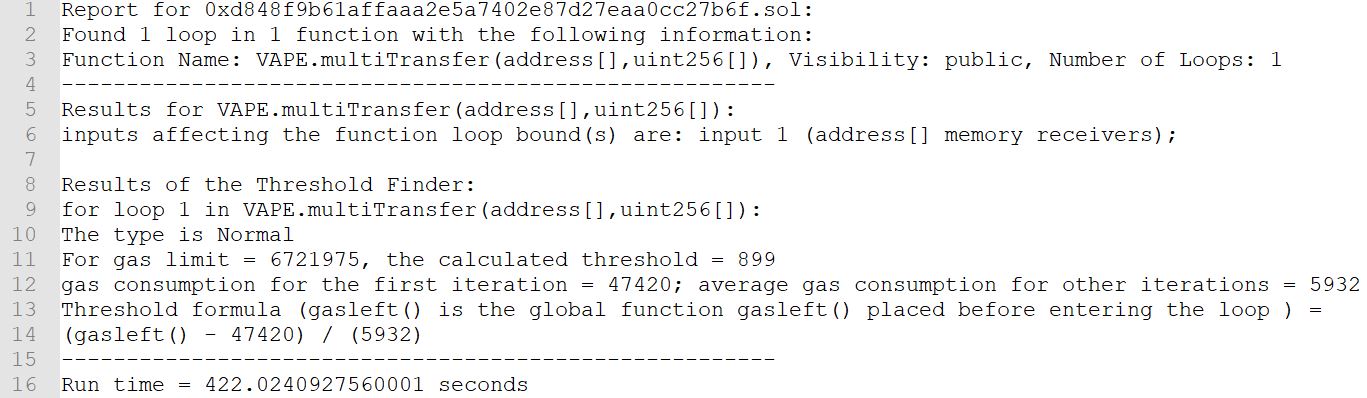}}
\centering
\caption{Output of \core}
\label{thr}
\end{figure*}

\subsubsection{Threshold of the Nested Loops}

The process for nested loops is slightly different. These loops and their order are identified by the Target Block Detector. Then the \core finds the threshold for the most inner loop and works its way back to the most outer loop. In order to find the threshold for the inner loops, the loop bounds for the outer loops are set to one, and the threshold for the target loops is found using the method mentioned before. The loop bound for the inner loop is set to zero, and the outer loop threshold is found. The output report contains a formula based on the threshold values of the outer and inner loops. The code snippets in Figure~\ref{nested1}, ~\ref{nested2} and ~\ref{nested3} demonstrate how a contract with nested loops are modified. The first code snippet shows the original function containing a nested loop. The second and third code snippets show the changes in the code in order to obtain the threshold for the inner and outer loops respectively.

\subsubsection{Output of the \core}

The output contains the signature of the functions containing loops and the number of loops in them.  It also has the variables affecting the loop bounds and their datatype. It provides the type of the loop (Normal/Single or Nested). Then, it provides the value of the threshold found by the tool for the provided gas limit along with the gas consumption of the first iteration and the average gas consumption for the other iterations. An example of the output of the \core\ is provided in Figure~\ref{thr}. In this example, the size of the input variable "receivers" is the bound of the loop,  the "require" statement looks like this:

require($ \text{receivers.length} < (\text{gasleft()} - 47420) / (5932)$,

"Loop bound is over the threshold!" );

In this case, if the user inputs an array of size greater than 899, the "require" statement gets triggered, and the execution stops before entering the loop. Finally, the threshold formula is given in the following format:
\[(gasleft() -gas\_1) / (gas\_2).\]

If there are nested loops in the contract, the formula is slightly different. The formula for the inner loops is similar to the ones above, but it considers the outer loop has only one iteration, and for the outer loops is similar to the following:

\[(gasleft() - gas\_1) / (gas\_2 + Internal).\]

Where \texttt{gasleft()} is the value returned by the function \texttt{gasleft()} placed right before the loop in the source code, \texttt{gas\_1} is the gas consumption of the first iteration of the loop, \texttt{gas\_2} is the average gas consumption of the other iterations, and \texttt{Internal} is the gas consumption of the internal loops.

This formula can be used in a \texttt{require} statement to prevent \OoG\ exceptions for that loop. The statement has to be placed in the source code right before the loop.  The user has to find the loop bound and place it in the \texttt{require} statement. So, an example of the \texttt{require} statement looks like this: 

\medskip
\noindent \texttt{require( $\text{loopBound} < (\text{gasleft()}- \text{gas\_1}) / (\text{gas\_2})$, "Over the limit")}
\medskip

Adding appropriate require statements can be a simple fix to many contracts containing user-controlled loops. However, this might not be the case when the contract uses poor implementation patterns. These \texttt{require} statements help the callers save some gas (by hitting the \texttt{require} statement rather than the block gas limit). Unfortunately, if the contract relies upon some user-controlled loops in order to function properly, other correction approaches are required, or the funds are still likely going to be locked. However, the loop bound information output by our tool can be very useful for the developers to come up with other solutions.

\section{Experimental Evaluation}
\label{evaluation}

\begin{table*}[ht]
\small
\newcommand{\tabincell}[2]{\begin{tabular}{@{}#1@{}}#2\end{tabular}}

\caption{\dete\  Comparison on \numcontracts\  Smart Contracts with loops.}
\label{table1}
\resizebox{\textwidth}{!}{
\begin{tabular}{|c|c|c|c|c|c|}
\hline
 \textbf{Tool Name} & \textbf{Method} & \tabincell{c}{\textbf{Number of Contracts}\\ \textbf{Successfully Scanned}}& \tabincell{c}{\textbf{False Negative}\\ \textbf{Rate(\%)}} & \tabincell{c}{\textbf{Average}\\ \textbf{Run-time (s)}}\\ 
\hline

\textbf{\tool} & \textbf{Static + Dynamic Analysis} & \textbf{997}  &\textbf{0} & \textbf{10} \\
\hline
GASTAP/Gasol & \tabincell{c}{Static Analysis } & 120 & 36  & - \\
\hline
Madmax & Static analysis &  921 & 79 &- \\
\hline
MPro & \tabincell{c}{Static Analysis +\\ Symbolic Execution} & 851 & 100 &242  \\
\hline
Mythril & Symbolic Execution & 870 & 100 & 3109  \\
\hline
Securify 2.0 & Static Analysis & 548  & 47 &176 \\
\hline
Slither & Static Analysis & 997 & 85 &2.6 \\
\hline
SmartCheck & Static Analysis & 1000 & 47 & 2\\
\hline

\end{tabular}
}
\end{table*}

\subsection{Real-world Smart Contract Benchmark}
We gathered a benchmark of \numcontracts\ real-world Solidity smart contracts containing over 60,000 functions from Etherscan\footnote{\url{https://etherscan.io}}. All contacts are gathered starting from the latest block at the time (Block Height: 11661369) and going backward in the chain. Each contract has 413 lines of code, 63 functions, 4 functions with loops, and 5 loops on average. These contracts were manually checked to ensure that each contained at least one loop. Also, information such as the number of functions, loops, and lines of code was manually collected. Hence, these values are used as the ground truth. \tool\ and the benchmark are available at \color{blue} \url{https://gasgauge.github.io/} \color{black}. The name of each contract file in the benchmark represents the contract address. 

\subsection{Experimental Setup}
We ran our experiments on a machine that was equipped with 8GB of RAM, a 4-core Intel Xeon 2.2 GHz processor with Ubuntu 18.04 running. To test \tool, the latest version (on January 2021) of nodejs, Ethereum, truffle, ganache-cli, solc-select, python3, and Slither were installed. We deployed the target smart contract to a test chain using Ganache-cli. Since \tool\ does not make any modifications to Truffle and Ganache-cli in its implementation, future versions of these tools are still compatible with \tool. In all these experiments, Solc compiler version 0.5.3 was used unless either the tool or the contract required a different version. Also, the block gas limit was so kept to the default value in Ganache-cli (6,721,975). The Solc compiler version and the block gas limit can be configured easily in the tool.

\subsection{Competing Tools}
The following seven tools were chosen to compare against \tool: GASTAP~\cite{Gastap}/Gasol~\cite{gasol}, Madmax~\cite{Madmax}, MPro~\cite{MPro}, Mythril~\cite{Mythril}, Securify 2.0~\cite{securify2}, Slither~\cite{Slither}, and SmartCheck~\cite{smartcheck}. MPro, Mythril, Securify 2.0, Slither, and SmartCheck were installed using the instruction provided on their official documentation. A combination of GASTAP and Gasol, which is available on a web interface \footnote{\url{https://costa.fdi.ucm.es/gastap}} was used. This interface was called using a script to scan our benchmark. MadMax is built into Contract Library\footnote{\url{https://contract-library.com}}. Therefore, we searched for our benchmark in Contract Library and collected the reports generated by MadMax. Thus, in our analysis, MadMax and GASTAP/Gasol do not have a run-time value associated with their effectiveness results. 

 \begin{figure*}[htp]
\centerline{\includegraphics[width= 6 in]{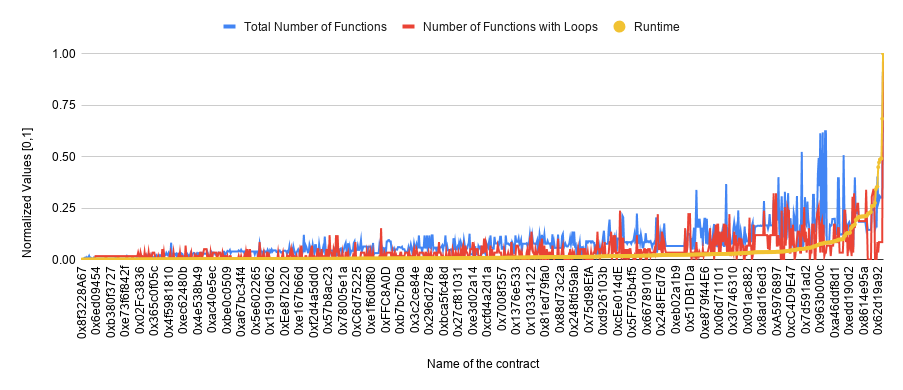}}
\centering
\caption{Effect of the Number of Functions on the Run-time}
\label{1}
\end{figure*}

 \begin{figure*}[htp]
\centerline{\includegraphics[width= 6 in]{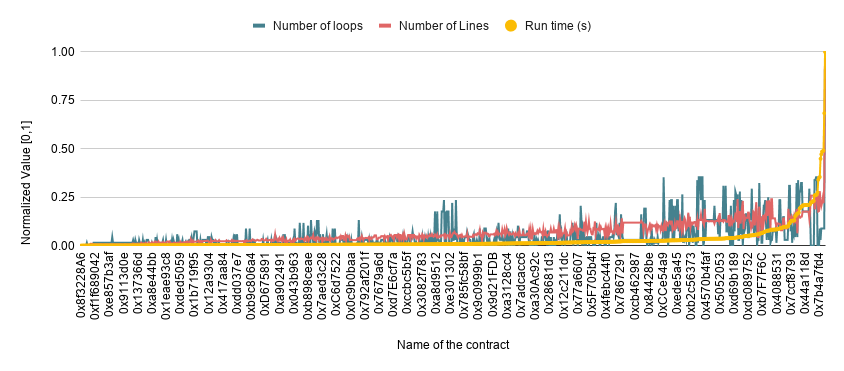}}
\centering
\caption{Effect of the Number of Loops and Line of Code on the Run-time}
\label{2}
\end{figure*}

 \begin{figure*}[htp]
\centerline{\includegraphics[width= 6 in]{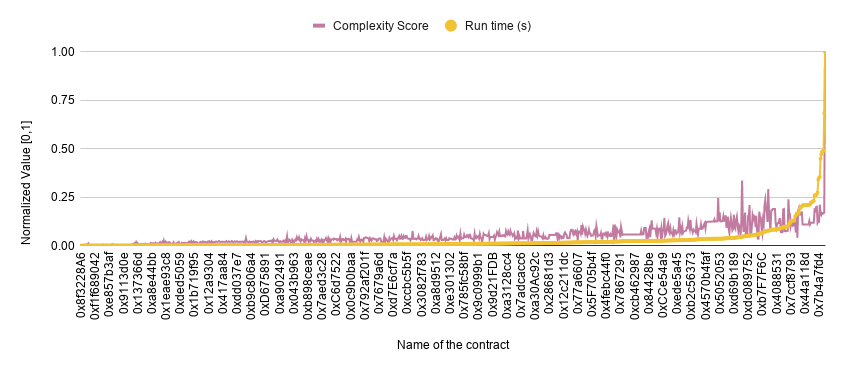}}
\centering
\caption{Effect of the Complexity Score on the Run-time}
\label{3}
\end{figure*}

\subsection{The Evaluation of the \dete}
This experiment aims to determine how accurate and efficient the \tool\ and the above-mentioned seven competing tools are at detecting potential loops for \OoG\ DoS vulnerabilities. All results are summarized in Table~\ref{table1}. The time-out limit for this experiment was set to 3 hours per contract for each tool. As can be seen, \tool\ was able to detect all the vulnerable loops in the contracts that it was able to scan. There were three contracts that Slither could not scan, and since the report generated by Slither is an essential part of our methods, our tool was not successful in checking them as well. Typically, a tool may not scan a contract for various reasons, like reaching the time-out and lacking support for the Pragma version. The false negative rate is calculated as the number of loops detected by the tool divided by the total number of loops in the contracts that the tool could scan. As mentioned before, the number of loops in each contract was manually collected and hence used as the ground truth. Also, we tested all tools on a benchmark of 1,000 contracts without loops, and they all had zero false positive rates. Hence, it is not listed in the table.

One can only obtain the reports of MadMax if a contract exists in Contract Library. Thus, "Number of Contracts Successfully Scanned" shows the number of contracts available on Contract Library. Thus, if a contract does not exist there, it is not necessarily a shortcoming of the tool. However, Madmax was not able to find \OoG DoS vulnerabilities in the Majority of the contracts, although it is one of the industry-standard tools for gas-related vulnerabilities. Also, although GASTAP/Gasol has the lowest false negative rate amongst the competitors, they have a low scan rate of around 12\%. The reason is that the Solidity compiler used by GASTAP/Gasol does not support a majority of the contracts in our benchmark or the generated report did not contain any meaningful information.

\tool\ has 0 false positive and false negative rates for the contracts supported by the tool and is efficient in scanning contracts (on average, about 10 seconds per contract). By contrast, all the other tools had difficulties detecting such vulnerabilities. Therefore, \tool\ is efficient and effective at detecting loops in smart contracts as it has an average run-time of 10 seconds and false negative and false positive rates of 0\%.

 \begin{figure*}[htp]
\centerline{\includegraphics[width= 6 in]{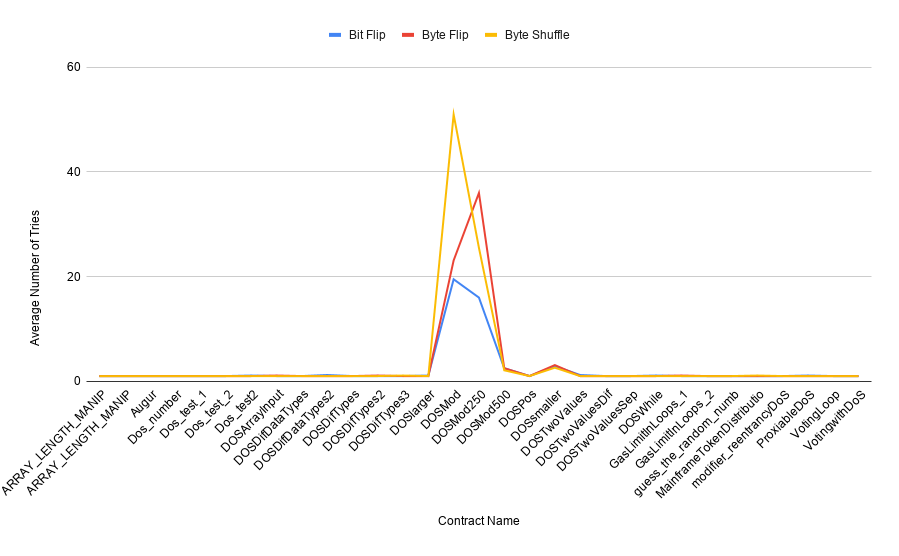}}
\centering
\caption{A Comparison of Different Methods for the White-box Fuzzer}
\label{fuzzer}
\end{figure*}

\subsection{Performance Analysis of the \dete}

In this experiment, we tried to find the main factors affecting the run-time of the \dete\ of \tool. We obtain the summary of each of the \numcontracts\ contracts containing loops in our benchmark. This summary contains the total number of functions, number of functions containing loops, number of code lines (only source code lines without comments and blank lines), and number of loops in each contract. We also measured the run-time for each of the contracts for the \dete. During this experiment, we concluded that the total number of functions and functions with loops have some impact on the run-time. However, these effects are not as noticeable as the number of loops and lines of code. It appears that the factors impacting the run-time of \dete\ in descending (run time) order are the number of lines of code, number of loops, number of functions containing loops, and total number of functions. This is also expected because \dete\ uses a static analysis approach as its primary method. 

ConsenSys has a tool called Solidity-metrics~\cite{metric}. This tool provides Source Code Metrics, Complexity, and Risk profile reports for projects written in Solidity. One of the factors they provide is "Complexity Score", a custom complexity score derived from code statements known to introduce code complexity such as branches, loops, calls, and external interfaces. We obtained this score for each of the contracts and observed the impact on the run-time. Based on our results, this factor has a noticeable correlation with the run-time. Therefore, the main factors in determining the run-time can be summarized into the "Complexity Score" of the code. Figures ~\ref{1}, and ~\ref{2} demonstrate the impact of different factors on the run-time of the \dete. These factors are the number of lines of code, number of loops, number of functions containing loops, and total number of functions. Figure ~\ref{3} demonstrates the impact of "Complexity Score" factor on the run-time of \dete.

This experiment cannot be done fairly on the other two phases since they contain run-time analysis. Thus, the time to deploy and run the contracts has a significant impact on the overall run-time of the tool. Furthermore,  for the \iden, factors like the search space, the ability of the fuzzer to find the right set of inputs, and the complexity of the target functions. Also, for the \core, the estimated value reported by the Threshold Estimator, cyclomatic complexity of the loop, and the actual threshold values are some of the factors that affect the run-time noticeably and cannot be measured easily.

\subsection{The Evaluation of the \iden}

In this experiment, we evaluated the results of the \iden. Generally, finding an \OoG\ instance requires a run-time analysis-based technique. To the best of our knowledge, \tool\ is the first tool, and the \iden\ is the first method that uses a run-time fuzzing technique to detect gas-related vulnerabilities and automatically identify \OoG\ instances. As a result, we did not find a direct competitor to compare the results of the \iden. Therefore, we manually checked each contract to obtain the number of functions satisfying the condition of the fuzzer and the input variables affecting the loops. Only 979 functions in 501 contracts met these criteria. Then, we ran our tool on the benchmark and obtained the results. Lastly, we verified the results manually. The fuzzer detected 968 functions in 499 contracts and identified their variables correctly in 53 seconds per contract on average. Two of the contracts were not scanned by Slither, so the fuzzer was not able to identify 11 functions. The fuzzer identified 614 instances of \OoG\ in 331 contracts. Although the fuzzer detected the rest, it could not find an \OoG\ instance in them for various reasons, like reaching the maximum number of tries (it was set to 10, but it is customizable), the function contained structs, or the code reverted. However, even when the fuzzer could not find an out of the gas instance, it still provided the function signatures with loops and variables affecting the loop bounds. Overall, the \iden\  identified all the satisfying functions that it could scan. It also was able to identify 614 instances of \OoG\ in 331 contracts.  

\subsection{Evaluation of the Methods for the \iden}

Three methods were considered when designing the white-box fuzzer. The first approach was a random bit flip, as described before. The second one was a random byte flip, which flips every bit in the byte starting from the randomly chosen bit. Lastly, a random byte shuffle was tested. In this approach, the fuzzer chooses a random bit to flip, and then all the bits in the byte starting from the chosen bit get randomly shuffled. During this experiment, we obtained the following results: 
\begin{enumerate}
     \item Because our implementation finds the exact functions and variables to fuzz, in most cases, all three methods can find the desired output within the first two tries. These are simple cases when one input variable is the loop bound, so the fuzzer has a high chance of finding the correct value for the value.
    \item In some cases, when the bound is more complex, the fuzzer takes about 3 or 4 tries to find a correct set of inputs using any of the approaches. An example of this is when the difference between the values of two of the inputs is the loop bound.
    \item There are also some cases that the fuzzer needs to try more numbers. An example of this is when "input mod 250" is the loop bound, and any number of loop iterations more than 240 triggers the \OoG\ condition. These contracts were the determining factors, and as shown, bit flip outperformed the other two methods. Hence, the bit flip approach was chosen in our design. 
\end{enumerate}

Figure ~\ref{fuzzer} shows a chart comparing the three methods on a benchmark of 28 contracts containing a total of 31 functions with loops. We tested each method ten times on each function and recorded the average number of tries for each method until finding a set of inputs that causes the \OoG\ exception. The number of tries means the number of different combinations of inputs before finding a satisfying set. This number was bounded to fifty in our experiments in order to halt the process promptly.

\subsection{The Evaluation of the \core}

In this experiment, we evaluated the results of the \core. Our benchmark has 4415 loops. \tool\ was able to find the thresholds for 932 loops in 467 contracts. 779 of these thresholds were calculated using run-time verification with a high rate of accuracy, and 153 of these thresholds were estimated with 95\% accuracy. A threshold is estimated if it is greater than 5,000 or for any reason, the run-time verification is not able to find the correct number. The average run-time for each loop was about 389 seconds, which is much faster than manual processing. To the best of our knowledge, \tool\ is the first tool ever attempted to find the loop upper bound limits in a smart contract. The closest tool to our work is GASTAP/Gasol. However, based on our experiments, the provided web interface does not support most of the contracts in our benchmark. Meanwhile, manually finding the thresholds is a tedious process. Therefore, to get an idea of the accuracy, we randomly picked 10 of the contracts and found the thresholds of their loops manually. Then, we compared the manual results with the ones generated by \tool. Based on our results, the calculated threshold is about 2\% lower than the actual threshold. We expect the calculated thresholds to be within 5\% of the actual values and usually lower than the actual values since our modifications introduce an insignificant extra gas usage.

\section{Limitations of \tool}
\label{limit}
If the Slither tool is not able to scan a contract or does not identify loops or data dependencies, then all three phases of \tool\ miss important information that lowers its accuracy. Also, the time limit and compilation problems of Truffle Suite may cause our tool not to operate properly. Functions containing structs or multi-dimensional arrays are not supported by the \core\ and \iden\. A struct is just a custom type that can be defined within a contract, and because it is different in each contract, it can be very challenging to automate the test generation for such a construct. However, they are supported in the \dete, which is the phase that competes with other tools. Moreover, the contracts should be self-contained and written in Solidity to be used by \tool. Thus, contracts with external calls to other contracts or written in other languages are not supported.

\section{Case Study}
\label{study}

We performed a case study to evaluate \tool\ in real-world applications. Quantstamp \cite{Quantstamp} is a leading verification company that evaluates smart contract projects for security-related issues and code quality. We collected the Quantstamp contract security certification of Airswap\footnote{Available at \url{https://certificate.quantstamp.com/full/airswap}}, a peer-to-peer trading network for Ethereum, to identify gas-related vulnerabilities in this project. Airswap is used for trading the USD equivalent of about 9200 ETH/week (over 20 million USD/week with current ETH value). Loop concerns due to gas usage have been reported in two contracts, \texttt{Swap.sol}, and \texttt{Index.sol}. \tool\ was able to identify one vulnerable function in \texttt{Swap.sol} and two vulnerable functions in \texttt{Index.sol}. \tool\ detected the loop and variables affecting the loop bound in \texttt{Swap.sol}. Although it did not find the threshold of the loop since the constructor of that contract takes a struct as an input, it detected the loops and variables affecting the loop bounds of  \texttt{Index.sol}. For the two other functions, it found the threshold of the loop as well. The threshold of one of these loops was over 5,000, so \tool\ estimated the threshold based on the average gas consumption of a few iterations.

Usually, to find the loop threshold values, one needs to audit the contracts to find potentially vulnerable code blocks. The next step is to examine these code blocks to understand what variables affect the loop bounds. Once all the information is gathered, a gas analysis needs to be performed. The gas analysis consists of making required files for tools like Truffle Suite to test a contract with different test cases to find the threshold for each loop\footnote{The process can be found at \href{https://github.com/airswap/airswap-protocols/issues/296}{here} and \href{https://github.com/airswap/airswap-protocols/pull/309/commits/9eb79d6af97f7428fdaa6d14224e57bf9b5d272d}{here}}. This process is tedious and requires many man-hours (estimated to be around 4 hours per contract) of work by the developers. However, \tool\ took about 10 minutes to find the vulnerable functions and loop thresholds. Also, \tool is automatic and requires limited resources and supervision. This case study shows that \tool is practical and can be used to save hours of manual work.

\section{Related Work}
\label{related}
Many smart contract verification tools scan a contract for multiple security vulnerabilities. Some of the most well-known tools are Manticore~\cite{manticore}, MPro~\cite{MPro}, Mythril~\cite{Mythril}, Securify~\cite{securify} (deprecated), Securify 2.0~\cite{securify2}, Slither~\cite{Slither}, SmartCheck \cite{smartcheck}, Verx~\cite{Verx}, and Zeus~\cite{Zeus}. Although these tools can detect multiple security vulnerabilities, most of them either cannot identify gas-related vulnerabilities or their results are not reliable due to their high false negative rate (See Table~\ref{table1}). Fuzzing tools like ContractFuzzer~\cite{contractfuzzer}, Echidna~\cite{echidna}, and Harvey~\cite{harvey} do not discover gas-related vulnerabilities, to the best of our knowledge. Meanwhile, some research has focused on gas-related vulnerabilities. GASTAP~\cite{Gastap} derives gas upper bounds for all public functions of smart contracts via inferring size relations, generating gas equations, and solving these equations. Madmax~\cite{Madmax} uses a static program analysis technique to detect gas-focused vulnerabilities automatically. However, most of these tools cannot detect gas-related DoS vulnerabilities directly, or they do not provide any information to fix the problem. 

\tool\ can find all the loops in a contract reliably and quickly. Also, it identifies the exact functions and variables and provides an \OoG\ instance that helps developers investigate the problem further. Finally, to the best of our knowledge, \tool\ is the only tool that accurately and reliably finds the threshold values and provides more useful information like each loop's type and the variables affecting it. This information is helpful in order to repairing the code and preventing gas-based attacks like DoS with Block Gas Limit. 

\section{Conclusions and Future Work}
\label{Conclusion}

Because smart contracts contain monetary transactions, it is crucial to make sure they are risk-free. This paper summarizes the design and implementation of \tool, an automatic tool that helps developers and contract owners identify DoS with Block Gas Limit vulnerability and repair their code. This tool contains three powerful sections. These sections use static analysis, run-time verification and white-box fuzzing to detect all the contract loops, provide an instance of out-of-gas and determine when the transaction starts to go out of gas. Finally, our experimental evaluation results on \numcontracts\ real-world Solidity smart contracts show that all the  methods are accurate and efficient. \tool\ only supports self-contained contracts, so contracts with external calls to other contracts are not supported. As a part of future work, the Mainnet forking will be used to include external contracts without source code. Also, The current implementation only supports contracts written in Solidity. As an improvement, we plan to extend our tool so that it can support more programming languages like Vyper and Rust. Furthermore, the fuzzer can be improved so it can identify the private functions containing loops and fuzz the public functions that call those private ones.

\bibliographystyle{IEEEtran}
\bibliography{Ref}
\end{document}